\documentclass[aps,pra,twocolumn,groupedaddress]{revtex4-1}

\usepackage{amsmath}
\usepackage{amsthm}

\usepackage{bbm,graphicx,bm,hyperref}   
\usepackage[pdftex]{color}

\def\R{R\'enyi }
\def\UXY{U_{\rm XY}}
\def\U3{U_{\rm D}}
\def\tS{\tilde{S}}
\def\rA{\rho_{\rm A}}

\def\ii{{\rm i}}
\def\sx#1{\sigma^{\rm x}_{#1}}
\def\sy#1{\sigma^{\rm y}_{#1}}
\def\sz#1{\sigma^{\rm z}_{#1}}

\def\Z2{\mathbbm{Z}_2}

\def\tr#1{{\rm tr}(#1)}
\def\1{\mathbbm{1}}

\def\ket#1{{| #1 \rangle}}
\def\bra#1{{\langle #1 |}}

\def\tit#1{{\em #1},}

\usepackage{booktabs}
\usepackage{multirow}
\AtBeginDocument{
\heavyrulewidth=.08em
\lightrulewidth=.05em
\cmidrulewidth=.03em
\belowrulesep=.65ex
\belowbottomsep=0pt
\aboverulesep=.4ex
\abovetopsep=0pt
\cmidrulesep=\doublerulesep
\cmidrulekern=.5em
\defaultaddspace=.5em
}

\begin{document}

\title{Entanglement growth in diffusive systems}

\author{Marko \v Znidari\v c}
\affiliation{Physics Department, Faculty of Mathematics and Physics, University of Ljubljana, 1000 Ljubljana, Slovenia}

\date{\today}

\begin{abstract}
We study the influence of conservation laws on entanglement growth. Focusing on systems with $U(1)$ symmetry, i.e., conservation of charge or magnetization, that exhibits diffusive dynamics, we theoretically predict the growth of entanglement, as quantified by the \R entropy, in lattice systems in any spatial dimension $d$ and for any local Hilbert space dimension $q$ (qudits). We find that the growth depends both on $d$ and $q$, and is in generic case first linear in time, similarly as for systems without any conservation laws. Exception to this rule are chains of 2-level systems where the dependence is a square-root of time at all times. Predictions are numerically verified by simulations of diffusive Clifford circuits with upto $\sim 10^5$ qubits. Such efficiently simulable circuits should be a useful tool for other many-body problems.
\end{abstract}





\maketitle

{\em \bf Introduction.--} 
Entanglement is one of crucial quantum resources responsible for the emerging 2nd quantum revolution -- exploiting quantumness to perform tasks not possible by classical means, for instance, quantum computation, teleportation, or secure communication~\cite{nielsen}. Even if not easily measurable~\cite{harvard}, it is an extremely powerful theoretical concept. This was further underlined by another discovery from '80, from a seemingly unrelated field, namely the quantum Hall effect~\cite{hall80}. It gradually brought to light the fact that there can be phases of matter that have topological order which goes beyond the Landau's paradigm of classifying all phases of matter just by local order parameters. Today we understand that such topological order is connected to certain patterns of entanglement~\cite{book}. A modern view in fact uses entanglement to distinguishing different phases of matter~\cite{wen11,schuh11}. Entanglement though plays a role also beyond the equilibrium phases. An example is for instance a putative non-thermal many-body-localized phase~\cite{mbl}, one of the distinguishing features of which is slow logarithmic-in-time growth of entanglement~\cite{MBLlog}.

Conservation laws and the associated symmetries are one of the most important properties of laws of physics. On the smallest scale, the elementary particles differ by their symmetries, and on the large scale, as well, the most violent objects we know -- black holes -- are believed to be defined only by their conserved quantities, charge, mass and angular momentum~\cite{nohair}. Furthermore, the symmetry to translations in time and its associated generator is the very object that governs dynamics. In short, symmetries are crucially responsible for the simplicity of nature at its core.

An important question is what role do conservation laws play on the dynamics of entanglement? Its growth with time is important also from a practical point of view. Namely, if it is small then efficient classical simulation of such systems is possible~\cite{vidal}. For generic local systems and initial states one expects that dynamics explores the whole available Hilbert space and therefore entanglement grows linearly with time. This holds true even for integrable systems, see e.g.\cite{calabrese2,alba17}. Because symmetries are about constraints, and because entanglement is given essentially by the number of degrees of freedom (two-level systems) involved, one might argue that symmetries will certainly affect entanglement growth. In the other hand, however, in the thermodynamic limit (TDL) one could also argue that conservation of a single charge should not matter much in a large Hilbert space. 

Therefore it was surprising and interesting when it was shown~\cite{pollmann} (focusing on diffusive 1D systems with conserved charge) that the entanglement, as quantified by the \R entropy 
\begin{equation}
S_r(t):=\frac{\log_2({\rm tr_A}\,{\rA^r})}{1-r},
\end{equation} 
grows in fact as $S_2 \sim \sqrt{t}$ starting from a generic separable initial state, instead of the ``expected'' $S_2 \sim t$, see also~\cite{huang19,zhou19}. This finding, if holding for generic systems, would have many consequences. For instance, one could argue that simple charge conservation causes the ``\R complexity'' $\sim 2^{S_2}$ to grow only as $\sim b^{\sqrt{t}}$, i.e. slower than exponentially (though still super-polynomially). A system with {\em diffusive conserved charge} would seem to be a less powerful quantum information resource than a one without it.

\begingroup
\squeezetable
\begin{table}[t!]
\begin{ruledtabular}
\begin{tabular}{cccccc}
$d$ & $q$ & $S_2(t<t_1)$ & $t_1$ & $S_2(t_1<t<t_\infty)$ & $t_\infty$ \\
\midrule
1D  & 2  & $c t$ & ${\cal O}(1)$  & $\sim c \sqrt{t}$ & ${\cal O}(L^2)$\\
2D & 2  & $\frac{l}{2L}t$ & ${\cal O}(L)$ & $\sim \sqrt{Lt}$  & ${\cal O}(L^3)$\\
3D  & 2 & $\frac{A}{3L^2}t$ & ${\cal O}(L^2)$ &  $\sim \sqrt{L^2t}$ & ${\cal O}(L^4)$  \\[2pt]
$d$ & 2 &  $\frac{A^{(d-1)}}{dL^{d-1}}t$ & ${\cal O}(L^{d-1})$ &  $\sim \sqrt{L^{d-1}t}$ & ${\cal O}(L^{d+1})$  \\ 
$d$ & $\ge 3$  & $\sim t$ & $t_1=t_\infty$ & & ${\cal O}(L^d)$ 
\end{tabular}
\end{ruledtabular}
\caption{Entanglement growth in diffusive lattice systems of linear size $L$ in $d$ spatial dimensions with $q$-level local Hilbert space (subsystem size is also $\propto L$). For qubits one has two regimes: linear growth for $t<t_1$, and a slow square-root growth at later times, before the finite-size saturation $S_2(t_\infty)\sim L^d$ is reached at $t_\infty$. For qudits the growth is linear. Time units are such that ${\rm d}S_2(0)/{\rm d}t \approx 1$ for all $L$ and $d$ (see text).}
\label{tab:summary}
\end{table}
\endgroup
We address the question of the \R entropy growth in local systems in any spatial dimension $d$ and for any local Hilbert space dimension $q$. Theoretical predictions for a bipartition ($L$ is the linear system size, being also proportional to the subsystem size), summarized in Table~\ref{tab:summary}, are numerically verified on large systems, with the total number of qubits up-to e.g. $252\times252\approx 6\cdot 10^4$ in 2D, and $48\times 48\times 48\approx 10^5$ in 3D. While we confirm Ref.~\cite{pollmann}, the main new and interesting finding is that in higher $d$ and $q$ the asymptotic $\sqrt{t}$ growth is in fact not what one will typically observe. While for diffusive qubit systems (spin-$1/2$, i.e. $q=2$) the asymptotic growth is still $\sim \sqrt{t}$, it starts in $d>1$ only at a time when the entropy already becomes extensively large, $S_2 \sim L^{d-1}$, in other words, in the TDL the $S_2$ grows linearly with time at any finite value of the entropy. For qudit systems ($q>2$), even in $d=1$, one expects instead a linear asymptotic growth, except in cases where the dynamics of all diagonal operators is diffusive. In this respect the often studied qubit systems in 1D are rather special -- diffusive growth is there observed already at early times and for {\em all diffusive systems} because a single diffusive charge, together with an identity operator, already exhausts the algebra of diagonal operators. As a side results, the presented new class of efficiently simulable systems with nontrivial dynamics could be useful in addressing other questions of many-body physics.

{\em \bf Theoretical prediction.--}
A class of systems that we study are lattice systems with local nearest-neighbor (n.-n.) interactions in $d$ spatial dimensions and with $q$-dimensional local Hilbert space, whose dynamics has a nontrivial conservation of the total particle number or the total spin in $z$-direction (i.e., a $U(1)$ symmetry). The dynamics of that conserved degree of freedom is assumed to be diffusive, while the rest of dynamics is generic (we exclude integrable systems). Specifically, the influence of possible non-$U(1)$ symmetries is left for future. In numerical demonstrations we also focus on Floquet systems in order to avoid having to deal with an additional conserved quantity (the energy). Linear dimension is denoted by $L$, and the total number of qudits by $n:=L^d$.

We shall discuss the entropy growth as quantified by the \R entropy $S_{r}$ (integer index $r>1$) starting from a pure product initial state. We prefer $S_r$ over von Neumann entropy $S_{r \to 1}$ due to its analytical simplicity. In generic systems all $S_r$, including $S_1$, are expected to behave in the same way, whereas for diffusive systems the $S_1$ (which we don't discuss) can behave differently~\cite{pollmann}. We will mostly focus on $S_2$ as a representative case of $S_{r > 1}$. We remark that sometimes $S_2$, rather than e.g. $S_1$, is the more relevant quantity~\cite{eisert19}, and is furthermore also easier to measure~\cite{zoller18}. Using a bipartition to regions A and B the reduced density operator is $\rA(t)={\rm tr}_{\rm B}{\ket{\psi(t)}\bra{\psi(t)}}$. The size of region A will be extensive, $|A| \sim L^d$, in order to avoid the effects of measure concentration that becomes prominent when the ratio of subsystem Hilbert space sizes $|B|/|A| \to \infty$. Specificaly, when that ratio grows the reduced $\rA(t)$ would for a typical state $\ket{\psi(t)}$ be increasingly closer to $\sim \mathbbm{1}$~\cite{Karol}. More precisely, tracing a random state over $|B| \gg |A|$ results in a spectrum of $\rA$ whose relative deviation from a flat one is $\sim q^{-(|B|-|A|)/2}$~\cite{jpa07}, and become negligible. Therefore, because we are interested in the influence of dynamics on $S_2$, and not simple kinematic effects of Hilbert space sizes, we require $|A| \to \infty$ in the TDL. For finite $|A|$ the saturation value of $S_2$ would also be finite, so that one could not unambiguously diferentiate betwen different powers in $S_2 \sim t^\alpha$. To facilitate comparison of different $d$ and $L$ we will measure $t$ in such units that one will generate a unit of entanglement in a unit of time, $S_2(1) \sim {\cal O}(1)$, i.e., in the language of quantum circuits $\sim {\cal O}(1)$ gates connecting regions $A$ and $B$ are applied per unit of time. Compared to local Hamiltonian evolution this means a rescaling of time by $L^{d-1}$. None of our conclusions depends on the choosen time-units, i.e., on the values of potential crossover times.

Let us first argue why and how conservation of magnetization (charge) matters for the {\em long-time behavior} of $S_2$. As we shall see, in the TDL $S_2$ is self-averaging (which is expected for generic, i.e., chaotic systems) and we will for simplicity focus on the purity $I(t):=2^{-S_2(t)}=\tr{\rA^2(t)}$. A non-rigorous intuitive meaning of the entropy is that it measures the effective number of the explored degrees of freedom needed to ``describe'' $\rA(t)$. For purity one can write $I \sim \frac{1}{N_{\rm eff}}$, where $N_{\rm eff} \sim 2^{L_{\rm eff}}$ is the effective Hilbert space size on which $\rA$ is supported, resulting in $S_2 \sim \log_2 N_{\rm eff} \sim L_{\rm eff}$. 

More quantitatively, the average purity $I$ over all computational initial states is
\begin{equation}
\bar{I}(t)=\frac{1}{q^n}\sum_{\vec{\mathbf{c}}} {\rm tr}[D^{(\vec{\mathbf{c}})}_{\rm A}(t)]^2,
\label{eq:I}
\end{equation}
where $D^{(c_k)}_k:=\ket{c_k}\bra{c_k}_k$ is a basis of diagonal matrices (projectors) with $c_k \in \mathbbm{Z}^q$ labeling the local computational state, and $D^{(\vec{\mathbf{c}})}_{\rm A}(t)={\rm tr}_{\rm B}\left[U^t D^{(c_1)}_1\otimes \cdots \otimes D^{(c_n)}_n U^{-t}\right]$. We see that at large times what matters is the spreading of the reduced diagonal operators $D^{(\vec{\mathbf{c}})}_{\rm A}(t)$, specifically their Hilbert-Schmidt norm. In particular, the average purity gets contribution from all possible products of initial projectors, where at each site we have $q$ different ones. The operator spreading in diffusive systems has a rich structure, having in general diffusive and ballistic features, see Ref.~\cite{vedika18} for details. Operators with a large initial overlap with the conserved charge will have a hydrodynamic (power-law) tail, as well as some other operators (a trivial example is the associated conserved current). Such operators will tend to cause diffusive behavior of $\bar{I}$. On the other hand, regardless of the diffusion, most operators will not exhibit any diffusive tails at long times. While in eq.(\ref{eq:I}) one actually needs the reduced $D^{(\vec{\mathbf{c}})}_{\rm A}(t)$, regardless of details one can say that if the dynamics of {\em all diagonal} operators is diffusive, one expects diffusive $\bar{I}$ and $S_2 \sim \sqrt{t}$. For qubits, $q=2$, there are just two local diagonal operators, $\1_k$ and $\sz{k}$, and therefore if $\sz{k}$ is diffusive one expects a long-time asymptotic growth $S_2 \sim \sqrt{t}$~\cite{pollmann}. However, for higher dimensional qudits, $q\ge 3$, the diagonal basis is spanned by $q$ linearly independent diagonal operators, only one of which is the conserved operator (the local magnetization). While one might think that diffusive modes that contribute to purity decay as ${\rm e}^{-\sqrt{t}}$ will due to their slow decay still dominate over non-diffusive ones, which decay as ${\rm e}^{-t}$, a simple counting argument shows that this is not to be expected. Namely, in a system of $n$ qubits the number of diagonal operators that are products of only diffusive magnetization $\sz{k}$ and the identity $\1_k$ is $2^n$, while the number of all other diagonal ones, that will in general be non-diffusive, is $(q^n-2^n)$. The diffusive contribution to $I$ will then be $\sim 2^n {\rm e}^{-\sqrt{t}}$ while a non-diffusive one is $\sim (q^n-2^n){\rm e}^{-t}$, so that in the TDL the non-diffusive contribution wins. Simply put, for higher $q$ there are exponentially more non-diffusive operators than diffusive ones. For generic $q\ge 3$ with only one conserved charge one therefore expects the asymptotic linear growth $S_2 \sim t$.

How about the {\em short-time behavior} of $S_2(t)$? We shall argue that it is, instead, always linear in time, even for qubits. Let us limit our discussion to qubits, $q=2$, as for $q\ge 3$ one anyway has linear growth $S_2 \sim t$ even at long times. For short-time behavior it is crucial to account for correlations spreading in a direction transversal to the boundary of dimension $d-1$ and area $A^{(d-1)}$ between regions A and B (in 2D $A^{(d-1)}$ is a circumference $l$, in 3D a true two-dimensional area $A$, in 1D the number of boundaries $c$, Table~\ref{tab:summary}). Starting from a product initial state the dynamics tries to generate entanglement across the boundary. For local (n.-n.) interaction the natural first candidate sites to be entangled are all $\sim L^{d-1}$ n.-n. pairs lying on the boundary between A and B. Only after all those qubits are entangled can a slowing down due to diffusion in a transversal direction kick in. Let us be more specific, with a view on numerical demonstration. In our random quantum circuits we will apply $L$ gates between random n.-n. qubits per unit of time. Such scaling is in-line with the mentioned units of time -- probability that such a random n.-n. gate connects A and B is $\sim \frac{L^{d-1}}{L^d}=\frac{1}{L}$, and therefore applying $L$ of them means we will have $\sim 1$ gates connecting regions A and B, and therefore, at least initially, generate one bit of entanglement in a unit of time. More precisely, the probability that a random gate connects A and B is $\frac{A^{(d-1)}}{dL^d}$, where the denominator $dL^d$ is the number of all nearest-neighbor bonds on a $d$ dimensional square lattice. The initial growth of entanglement is therefore expected to be
\begin{equation}
S_2 \approx \frac{A^{(d-1)}}{dL^{d-1}} t.
\label{eq:Sshort}
\end{equation}
We expect this linear growth to hold for any $S_r$, including $r=1$. Such linear growth will continue until the time $t_1 \sim L^{d-1}$ at which $S_2(t_1) \sim A^{(d-1)}$. After that one will crossover into the asymptotic diffusive growth $S_2 \sim \sqrt{t}$, until at $t_\infty \sim L^{d+1}$ a finite-size saturation value $S_2(t_\infty)\sim L^d$ is reached.

We see that in higher spatial dimensions the region of diffusive growth is parametrically small, it lasts from $t_1 \sim L^{d-1}$ till $t_\infty \sim L^{d+1}$. Furthermore, in the TDL it is pushed to infinitely large values of entropy $L^{d-1} \lesssim S_2 \lesssim L^d$ and will be hard to observe. Qubits in $d=1$ are rather special because the linear growth ends at $S_2 \sim L^0=1$ (i.e., at short time $t_1 \sim 1$) and one gets $S_2 \sim \sqrt{t}$ in the whole range of $S_2$ (and $t$). In short, in $d=1$ the asymptotic $\sim \sqrt{t}$ growth is ``easy'' to observe, while in $d>1$ it is hard because it appears in the TDL only at infinitely large values of $S_2$. Therefore the generic behavior after a quench from a product state is in $d>1$ the linear growth (which is as fast as allowed by the Lieb-Robinson bound~\cite{bravyi16}). Table~\ref{tab:summary} summarizes these findings. 
\begin{figure}[t!]
\centerline{\includegraphics[width=.75\columnwidth]{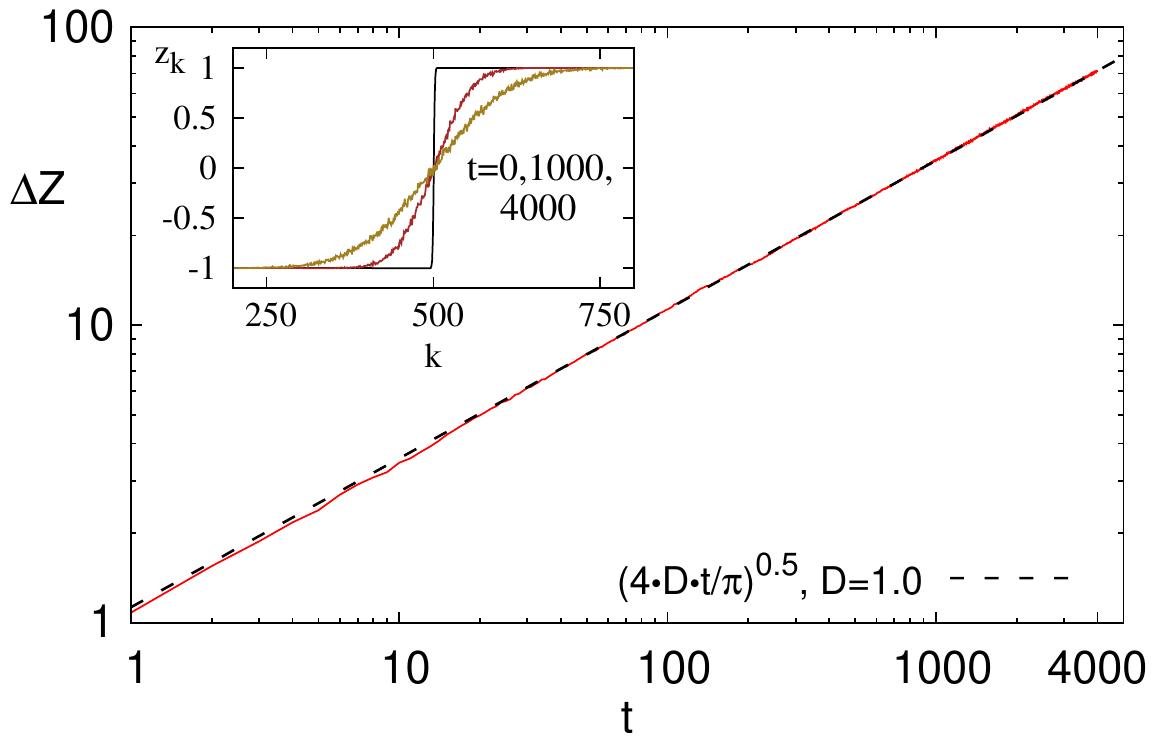}}
\caption{(Color online) Diffusive melting of a domain wall in a 1D random Clifford circuit with $\UXY$ gate and $L=1000$ sites. The inset shows the domain wall profile, while the main plot show a diffusive growth of transferred magnetization across the domain wall (averaged over $10^3$ circuit realizations).}
\label{fig:difuzija}
\end{figure}

\begin{figure*}[t!]
\centerline{
\includegraphics[width=.65\columnwidth]{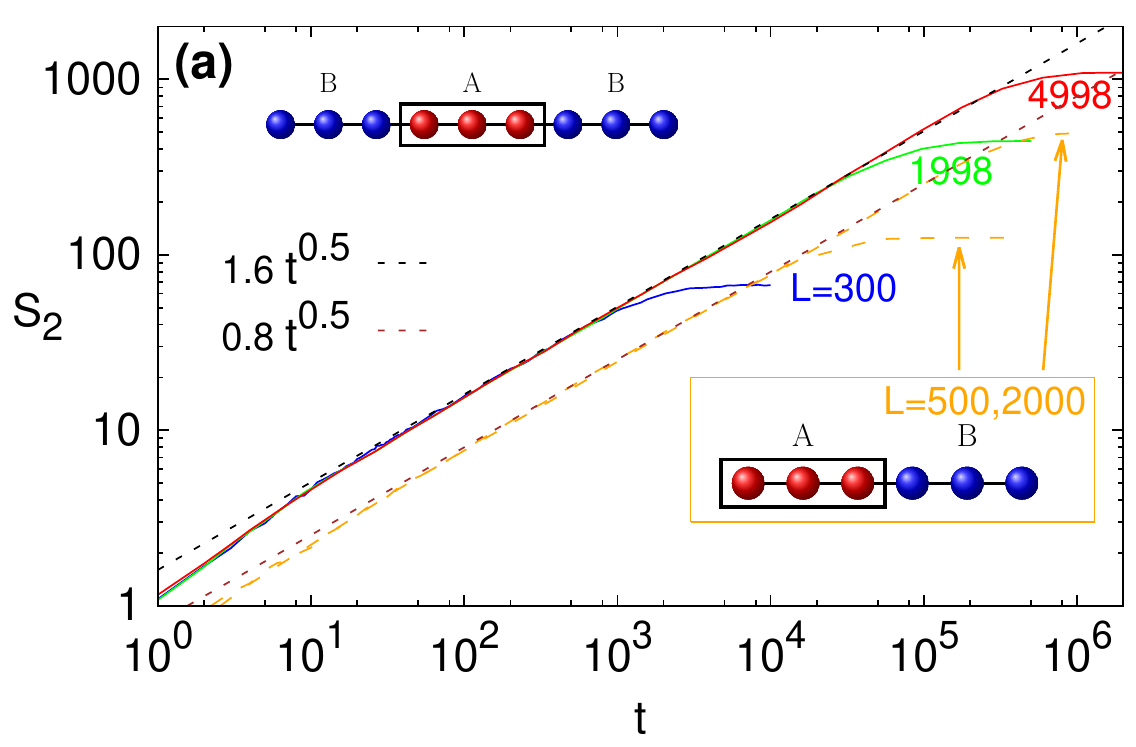}
\includegraphics[width=.65\columnwidth]{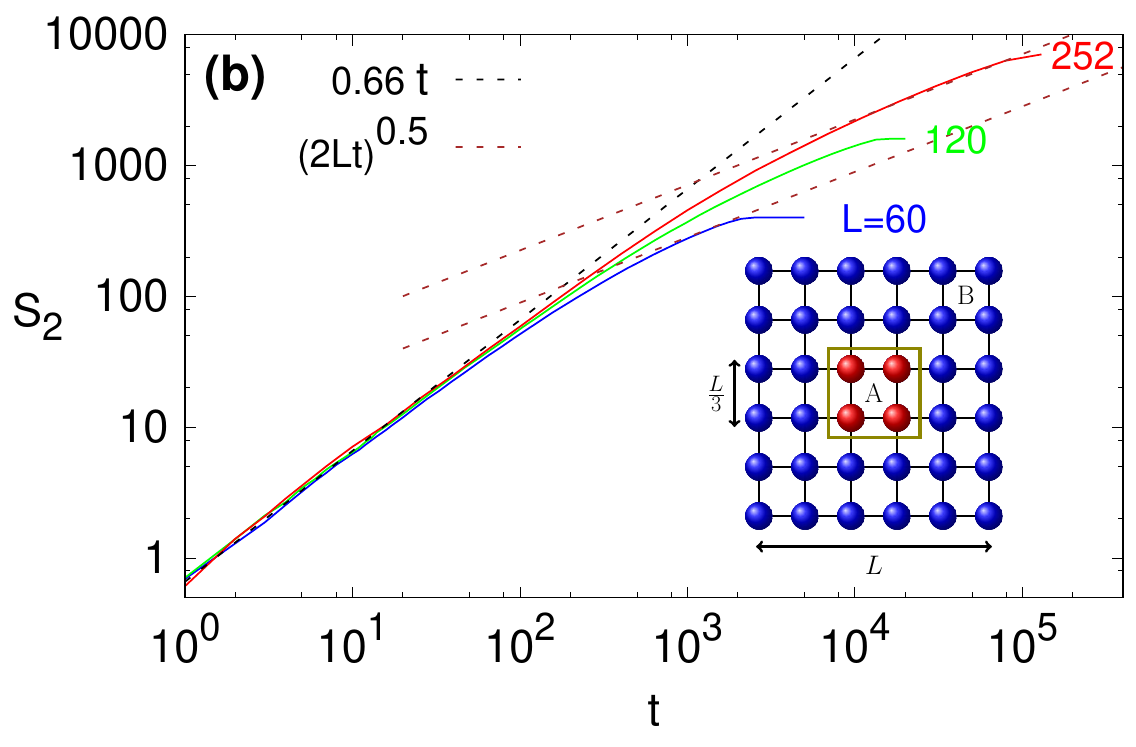}
\includegraphics[width=.65\columnwidth]{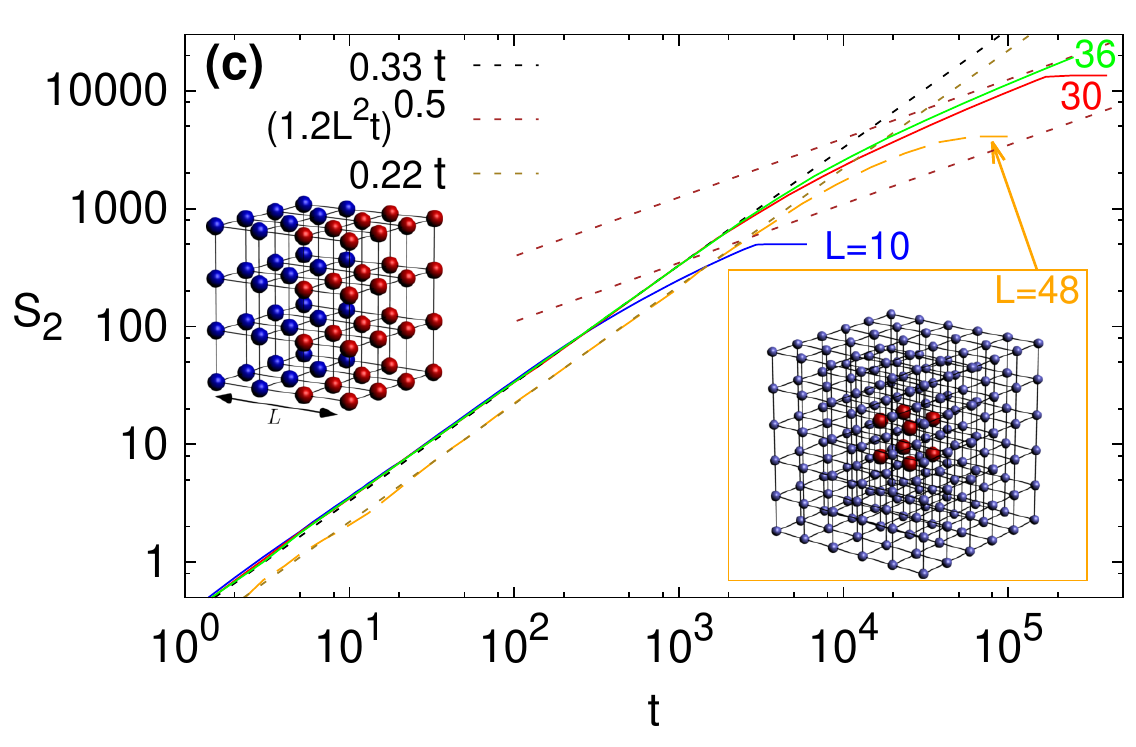}
}
\caption{(Color online) Entanglement growth in diffusive Clifford circuits in 1D (a) (solid curves are for the middle-$\frac{1}{3}$ bipartition), 2D (b), and 3D (c) (solid blue, red and green curves are for the half-cut bipartition).}
\label{fig:plot123D}
\end{figure*}
{\em \bf Clifford circuits.--} It is always useful to take the simplest model, analytically or numerically, that displays the physics one wants to explore. A setting for which one can get exact results for the entanglement dynamics are so-called random quantum circuits~\cite{emerson03} composed of a series of (random) local unitaries. Random circuits can be thought of as handy toy models of many-body physics but also as a useful theoretical concept called a unitary designs~\cite{brandao16}. One of the first exact results was obtained by rewriting the dynamics of purity on average as a classical Markov process~\cite{oliveira07}, mapping it to a solvable quantum spin chain and getting an exact expression for the gap $\Delta$ or the decay rate~\cite{exact}, i.e., entanglement speed~\cite{adam17} $v_{\rm E}$ in modern language. For instance, for a circuit composed of a random 2-site unitaries applied to a random n.-n. pair of qubits in a chain with $L$ sites, one gets~\cite{exact} $v_{\rm E}=(1-\frac{4}{5}\cos{\frac{\pi}{L}}) \asymp \frac{1}{5}$. If one would instead take a regular brick-wall pattern of applied gates, like in~\cite{pollmann18}, one instead has to calculate the gap of a product of Markovian matrices, obtaining a ``multiplicative'' form $v_{\rm E}=2\ln{\frac{5}{4\cos{\frac{\pi}{L}}}}$, going in the TDL to $v_{\rm E}\asymp 2\ln{\frac{5}{4}}$, as also calculated in~\cite{pollmann18,nahum18}. Studies of random circuits have expanded in recent years, including $U(1)$ conserving ones~\cite{vedika18}, with many nice exact results, see e.g.~\cite{adam17,nahum18,vedika18,pollmann18,chan18,chalker18,you19}. They have been also notably used in a race towards quantum supremacy~\cite{google}.

Let us check the above predictions for $S_2$ by numerical simulations of random circuits. In order to be able to simulate large systems we resort to the so-called Clifford circuits. For a $q$ level system the local generalized Pauli operators $X$ and $Z$ are defined~\cite{knill96,9802007} as
\begin{equation}
  X\ket{j}=\ket{j\oplus 1},\quad Z\ket{j}=\omega^j \ket{j},\quad j=0,\ldots,q-1,
\label{eq:XZ}
\end{equation}
where $\omega:={\rm e}^{2\pi\ii/q}$, and all additions are modulo $q$ (the sign $\oplus$). Generators of the local Pauli group are all $q^2$ products $X^vZ^w$, with $v,w=0,\ldots,q-1$. The generalized Pauli group (GPG) on $n$ sites is then formed by the tensor product of $q^2$ local Paulis, allowing also for all overall phases $\omega^j$. Due to $ZX=\omega XZ$, a product of two members of the GPG is again in the GPG. The action of such Pauli operators on the computational basis states is simple, for instance, $X^{x_1} Z^{z_1} \otimes \cdots \otimes X^{x_n}Z^{z_n}\ket{\bf a}=\omega^{{\bf z}\cdot{\bf x}}\ket{{\bf a \oplus x}}$.

Evolution of states is however not done by updating each computational basis state -- that would be inefficient for highly entangled states -- but rather by a stabilizer formalism~\cite{gottesman97}. A state $\ket{\psi}$ on $n$ qubits is called a stabilizer state if it is a unique joint eigenstate with eigenvalue $1$ of $n$ independent stabilizer generators $g_j$ from the GPG. For qubits one can obtain it as a product of projectors, $\ket{\psi}\bra{\psi}=\Pi_j (\1+g_j)/2$, for qutrits one has $\ket{\psi}\bra{\psi}=\Pi_j (\1+g_j+g_j^2)/3$. A Clifford circuit is a series of Clifford gates $U$, each of which preserves the GPG. That is, $U_{j,k}$ acting nontrivially on sites $j$ and $k$ maps a member of the GPG to another member of the GPG (instead of to a superposition of GPGs as for generic $U$). And here lies the advantage of Clifford circuits. Instead of updating the state $\ket{\psi}$ one instead updates each generator $g_j$, whose number is always $n$ and which will remain elements of the GPG~\cite{knill96,gottesman97}. Performing one gate, i.e., updating all stabilizers, takes ${\cal O}(n^2)$ operations. Entanglement and state overlaps can also be calculated efficiently~\cite{aaronson,audenaert05}.

A common choice of Clifford gates are the phase gate ${\rm P}\ket{j}=\omega^{j(j-1)/2}\ket{j}$, the Hadamard gate ${\rm H}\ket{j}=\frac{1}{\sqrt{q}}\sum_k \omega^{kj}\ket{k}$, and a 2-qudit controlled-NOT gate ${\rm CNOT}_{12}\ket{j,k}=\ket{j,k \oplus j}$. The dynamics of Clifford circuits therefore boils down to modular arithmetic~\cite{deMoor}. They also form a unitary 2-design~\cite{2design} (correctly reproduce Haar averages over all 2nd order polynomials in $\rho_{\rm A}(t \to \infty)$, e.g., a purity), with the same convergence behavior as generic random circuits. Therefore, in absense of conservation laws $S_2(t)$ behaves similarly for Clifford as well as for generic random circuits. So far Clifford circuits have been extensively studied in quantum information, but not so much in condensed matter or statistical physics. The reason being that their dynamics is typically either ballistic or localized~\cite{chris} (fluctuations though can exhibit interesting behavior~\cite{adam17}). We shall study a new class of random Clifford circuits that conserve magnetization and whose dynamics is diffusive. By looking at random circuits we are also able to focus exclusively on the role of the $U(1)$ symmetry without any stray effects caused by other conservation laws (e.g., conservation of energy).

{\em \bf Numerical verification.--}
Let us first focus on qubits. For qubits the elements of the local GPG are just the ordinary Pauli matrices $\{\sx , \sy , \sz , \1\}$. To preserve the total magnetization our Clifford circuit consists of applying the XY gate $\UXY:=\exp{(-\ii \frac{\pi}{4}(\sx{j}\sx{k}+\sy{j}\sy{k}))}$ to a randomly selected n.-n. pair of sites on a $d$ dimensional square lattice. It is easy to verify that $\UXY^\dagger \1_j\sz{k} \UXY = \sz{j}\1_k$ and $\UXY^\dagger \sz{j}\1_k \UXY = \1_j \sz{k}$, and therefore the total magnetization $\sz{j}+\sz{k}$ is conserved. It also implies that a pair of oppositely polarized spins is exchanged, $\UXY\ket{\!\uparrow\downarrow}=\ket{\!\downarrow\uparrow}$. Because the pair $(j,k)$ is chosen at random, it is also immediately clear that the dynamics of magnetization is diffusive, e.g., starting from a domain wall initial state $\ket{\!\downarrow\ldots \downarrow \uparrow \ldots \uparrow}$ the average profile at time $t$ can be expressed exactly in terms of binomial probabilities, that can be approximated in the large-$t$ limit by the error function (see Fig.~\ref{fig:difuzija} for an explicit numerical demonstration).

Starting with the initial state $\ket{\psi} \sim (\ket{\!\uparrow}+\ket{\!\downarrow})^{\otimes n}$ stabilized by $g_j=\sx{j}$, we can simulate our Clifford circuit for thousands of qubits up-to very long times, despite the entanglement eventually being a volume-law $S_2 \sim L^d$. Entanglement calculation is simplified by the fact that the state at any time is composed of an integer number $M$ of generalized EPR pairs, $\frac{1}{\sqrt{q}}\sum_j \ket{j}_1\otimes \ket{j}_2$, stabilized by two generators $X_1X_2$ and $Z_1 Z_2^{-1}$. Eigenvalues of the reduced density operator $\rA(t)$ are all equal, and using the base-$q$ logarithm one has $S_r=M$ for all $r$. In this respect Clifford circuits are special, however, their dynamics of $S_2$ will be generic and consistent with the presented theory. In addition, we will also numerically demonstrate that a similar behavior is obtained also for non-Clifford circuits where the spectrum of the reduced density operator is not flat. Therefore, while one can not extract the difference between different $S_r$ from Clifford circuit simulations (in particular that $S_1$ can behave differently), or use any finer measures of complexity that involve the individual spectral components, e.g.~\cite{Chamon14}, we argue that they do result in generic behavior of $S_2$. This is in-line with the fact that while Clifford circuits are not universal, already very small modifications, see e.g. Refs.~\cite{Zhou19,Eisert20}) (that might not influence many quantities), do result in universal behavior, e.g. universal quantum computation. For another solvable evolution that also results in a flat spectrum of $\rA$ see Ref.~\cite{prosen}. 

\begin{figure}[t!]
\centerline{\includegraphics[width=.95\columnwidth]{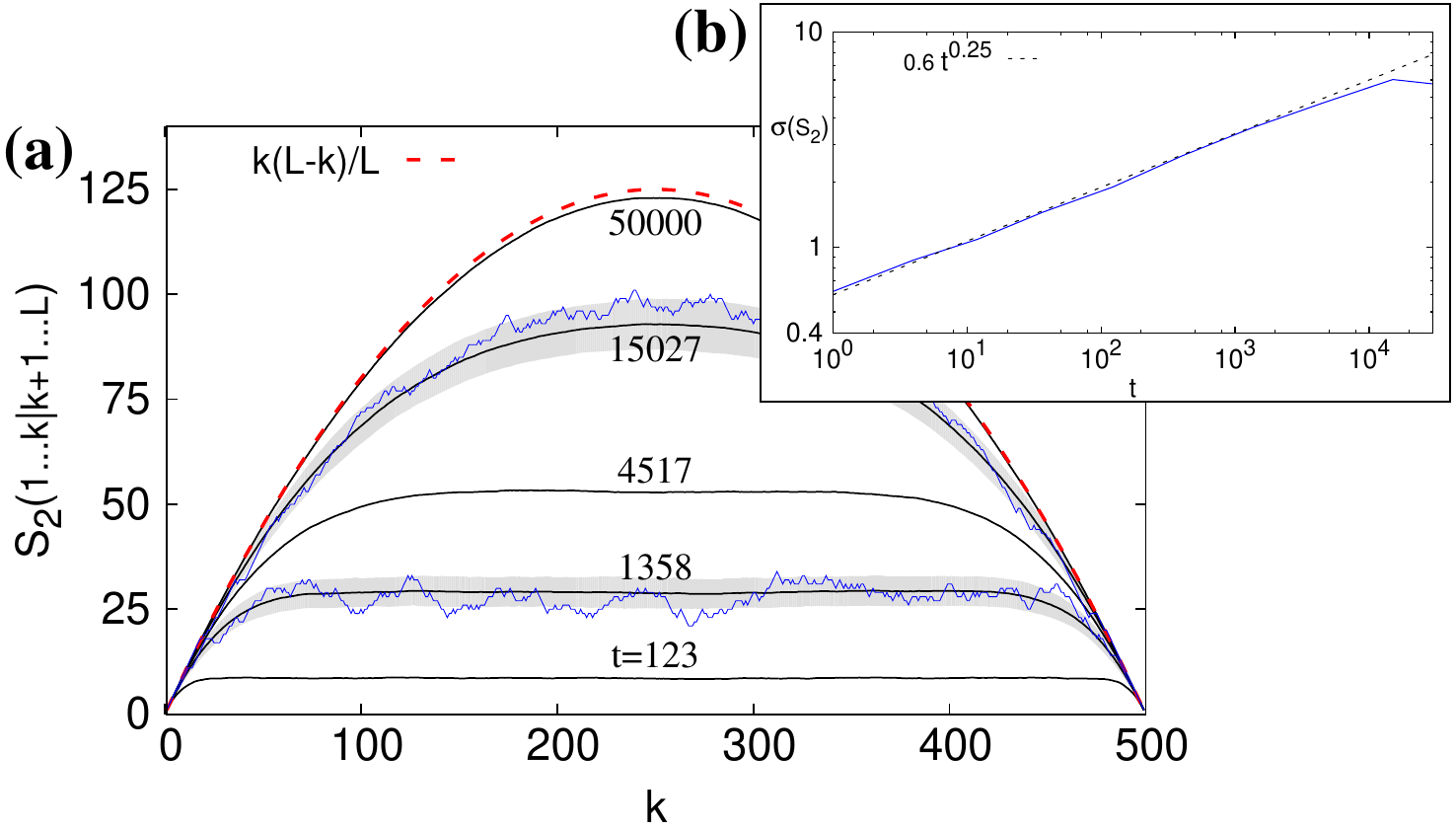}}
\caption{(Color online) Entanglement profile for a bipartite cut after the first $k$ sites (a), and fluctuations (b), in diffusive 1D Clifford circuit with $L=500$ (same data as in Fig.~\ref{fig:plot123D}(a)). In (a) we also show standard deviation (grey shading) and one realization at two selected times.}
\label{fig:flukt}
\end{figure}

In Fig.~\ref{fig:plot123D} we show $S_2$ for 1D, 2D, and 3D lattice, and for different bipartite splitting of $n$ spins into regions A and B. For 1D we see that the asymptotic growth is $S_2 = c \sqrt{2t/\pi}$ with $c$ being the number of boundaries between A and B ($c=1$ for a half-cut, and $c=2$ for the middle-$\frac{1}{3}$ cut). We also observe that at short times $t \lesssim 10$ the growth is a bit faster than diffusive. This means that in small systems $L \sim 30$ (being a typical maximal size amenable to other methods) it would be very difficult to see the true asymptotic growth over a significant range of times. In 2D we show data only for the case where the region A is the middle-$\frac{1}{3}$ part of the full square lattice with $n=L\times L$ qubits, as this bipartition gives a clearer transition between linear and diffusive growth (see Appendix~\ref{app:2D} for the half-cut data). Numerics confirms the short-time growth given by Eq.(\ref{eq:Sshort}) without any additional prefactors ($A^{(1)}=4L/3$). We also note that to see the asymptotic growth $\sim \sqrt{Lt}$ one needs fairly large systems; even for $L=252$ one can see only about one decade in time of $S_2 \sim \sqrt{t}$, while on the other hand three decades of $S_2 \sim t$. In 3D the situation is even less favorable for slow asymptotic diffusive growth. Nevertheless, in the more favorable half-cut bipartition we can see a transition from the short-time $S_2 = \frac{A^{(2)}}{3L^2}t$ to the long-time $S_2 \sim \sqrt{L^2 t}$. Again, for $t<t_1$ the linear growth (\ref{eq:Sshort}) has no additional prefactors (for a half-cut $A^{(2)}=L^2$, for the middle-$\frac{1}{3}$ cut $A^{(2)}=\frac{2}{3}L^2$). For the middle-$\frac{1}{3}$ cut, despite a large number of qubits, $n=48^3=110592$, one can barely hint the eventual $\sim \sqrt{t}$ growth. Finally, we show entanglement profiles (Fig.~\ref{fig:flukt}), i.e. $S_2$ for a bipartite cut with region A being the first $k$ spins. Also shown are the fluctuations of $S_2$ between different circuit realizations, showing that the relative fluctuations scale as $\sigma(S_2)/S_2 \sim 1/\sqrt{S_2}$, and therefore in the TDL at large times dynamics is self-averaging. It suffices to look at a single random circuit realization. In Appendix~\ref{app:2D} we also show data fore more complicated Clifford gates than $\UXY$, leading to similar results.

We also check the case of qudits with $q>2$, also studied in~\cite{pollmann}. To that end we simulate a qutrit chain ($q=3$, i.e. spin-$1$ particles) where the local diagonal basis is spanned by $\{ Z_k, Z_k^2=Z_k^{-1},Z_k^3=\1 \}$, and we take the initial state stabilized by generators $g_j=X_j$. Taking a Clifford circuit with the n.-n. gate being $\U3={\rm H}_2\, {\rm CNOT}_{21} {\rm CNOT}_{12}{\rm CNOT}_{12}\,{\rm H}_1$, which gives rise to diffusive conservative dynamics of both diagonal matrices, $\U3^\dagger \1_j Z_k \U3= Z_j \1_k$, $\U3^\dagger Z_j\1_k \U3= \1_j Z_k$, and $\U3^\dagger \1_j Z^2_k \U3= Z^2_j \1_k$, $\U3^\dagger Z^2_j\1_k \U3= \1_j Z^2_k$, we observe the expected diffusive $S_2 \sim \sqrt{t}$ (Fig.~\ref{fig:plotd3}). For further qutrit numerics see Appendix~\ref{app:q3}, including circuits with non-Clifford gates.
\begin{figure}[t!]
\centerline{\includegraphics[width=.8\columnwidth]{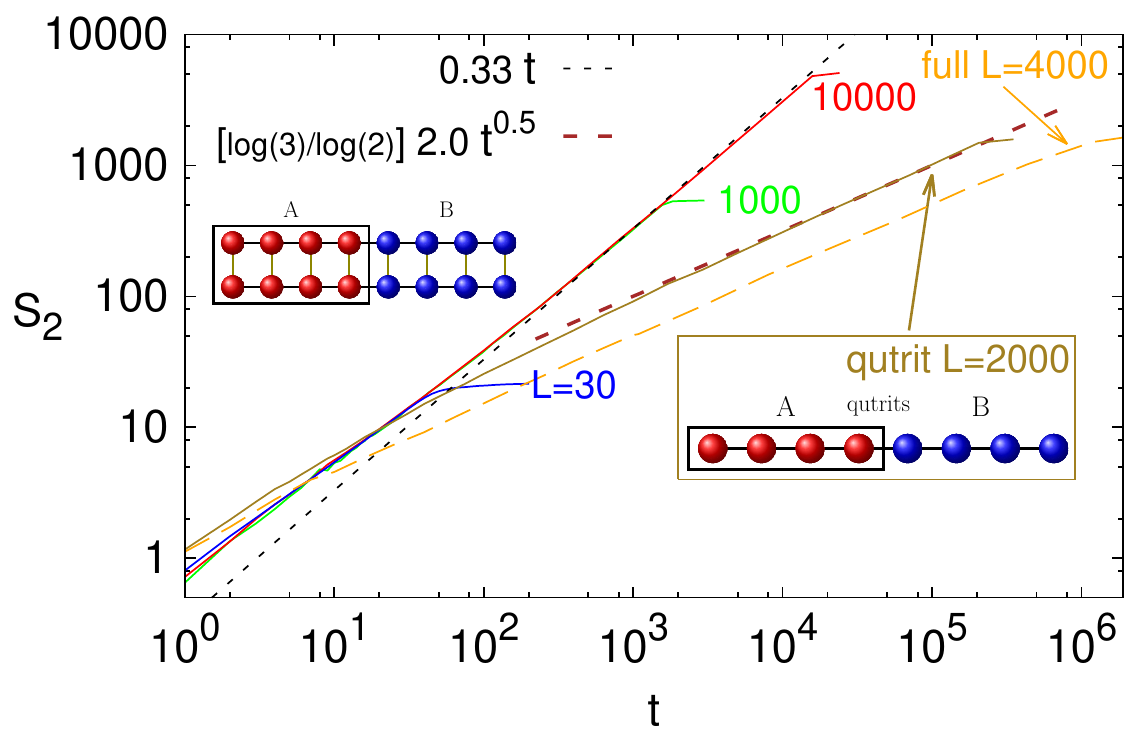}}
\caption{(Color online) Entanglement growth for 1D qudit systems with $q\ge 3$. Solid curves (blue, green, red) are for the ladder ($q=4$) with diffusive dynamics only on the upper leg, while the dashed orange curve (labeled ``full $L=4000$'') is for the ladder with $U(1)$ conservation on both legs. Solid olive curve is for the qutrit ($q=3$) chain with diffusive dynamics of all diagonal operators.}
\label{fig:plotd3}
\end{figure}

To get a generic case in which not all diagonal operators are conserved we use local dimension $q=4$ visualized as a qubit ladder ($n=2L$) with the local space corresponding to one rung. Per unit of time we apply $L$ steps, each consisting of: the gate $U_{\rm ZZ}=\ii \exp{(-\ii \frac{\pi}{2}\sz{1}\tau^{\rm z}_2)}$ applied to a random rung ($\sigma^\alpha$ are Pauli matrices on the upper and $\tau^\alpha$ on the lower leg), and either magnetization-preserving $\UXY$ on a random bond in the upper leg, or a non-conserving $U_{\rm G}={\rm CNOT}_{12}\,{\rm H}_1\,\exp{(\ii \frac{\pi}{4}\sz{2})}$ on a random bond of the lower leg. Magnetization is conserved only in the upper leg and one therefore expects generic behavior with $S_2 \sim t$. This is indeed observed in Fig.~\ref{fig:plotd3}. We can also see that for times less than $\approx 10^2$ slower growth is observed in which diffusive dynamics competes with increasingly dominating non-diffusive dynamics of other operators, and therefore, once again, one needs large systems with $L \gtrsim 100$ in order to see the true linear asymptotic growth. We contrast this linear growth with a special case: if we instead of $U_{\rm G}$ apply $U_{\rm XY}$ also on the lower leg, such that magnetization on both legs is conserved, one again gets a non-generic $S_2 \sim \sqrt{t}$ (orange dashed curve in Fig.~\ref{fig:plotd3}). We remark that this latter case of using $U_{\rm XY}$ on both legs corresponds to a Trotterized dynamics of the Hubbard chain (using Jordan-Wigner transformation the upper leg represents spin-up fermions, the lower spin-down, $U_{\rm XY}$ is hopping, while $U_{\rm ZZ}$ is the on-site interaction). In Appendix~\ref{app:q4} we show further ladder examples.

{\em \bf Conclusion.--}
We have presented a theory of the \R entropy growth in lattice systems that conserve the total magnetization due to $U(1)$ symmetry. We show that in general qubit systems the entanglement grows linearly in time until at an area-law value of $S_2$ a crossover to slower square-root growth happens. In 1D qubit systems the diffusive $\sqrt{t}$ growth is generic because the crossover happens already at small values of entanglement $S_2 \sim 1$, while in higher dimensions the regime of such growth is in the thermodynamic limit pushed to infinitely large values of $S_2$ (at any finite value of $S_2$ the growth is linear). For lattice systems with more than $2$ local levels (spin-$s$ particles, $s \ge 1$) and a single conserved charge non-diffusive degrees dominate and one expects the linear growth, irrespective of diffusion, an exception being a situation where the dynamics of all diagonal operators is diffusive. Two-level systems in 1D are special because a single diffusive charge exhausts all non-trivial local projectors. Entanglement growth can therefore distinguish both the spatial dimensionality as well as the size of the local Hilbert space, with the influence of a diffusive charge diminishing when either of the two increases.

It would be interesting to generalize our results to other transport types beyond diffusion, an abvious conjecture in 1D is that asymptotically one will have $S_{r>1} \sim t^{1/z}$, with $z$ being a dynamical transport exponent. An interesting question is the influence of more complicated symmetries than $U(1)$, as well as the presence of multiple symmetries. Energy conservation (i.e., time-translation symmetry) which comes automatically in autonomous Hamiltonian systems, and which was not discussed specifically, is an important example. While diffusive energy transport likely plays a similar role, it carries few technical complications, for instance, it is a nontrivial problem to establish diffusion of energy in the first place. A class of systems not touched uppon are integrable systems where one has an extensive set of local conserved quantities. Those will typically be ballistic, however it needs not be so. The question is can such non-ballistic modes influence the entropy growth in some generic fashion. In the present work we focus on $S_2$, and conjecture that all integer $S_{r>1}$ behave similarly, there could however be non-trivial time-scales connected with the index $r$. von Neumann entropy $S_1$ is special~\cite{pollmann}, and one could more generally study how the whole average eigenvalue spectrum converges to that of a random state~\cite{jpa07}. 

A promising direction is also employing introduced nontrivial Clifford circuits to further explore the many-body physics, and, more generaly, to understand their dynamical properties and in which aspects are they different than those of the many-body Hamiltonian systems.

I would like to thank C.~von Keyserlingk, F.~Pollmann, and T.~Rakovszky for comments on the manuscript, and acknowledge support by Grants No.~J1-1698 and No.~P1-0402 from the Slovenian Research Agency.

\newpage
\clearpage
\appendix

\section{Additional data for 1D and 2D Clifford systems}
\label{app:2D}

\begin{figure}[t!]
\centerline{\includegraphics[width=.8\columnwidth]{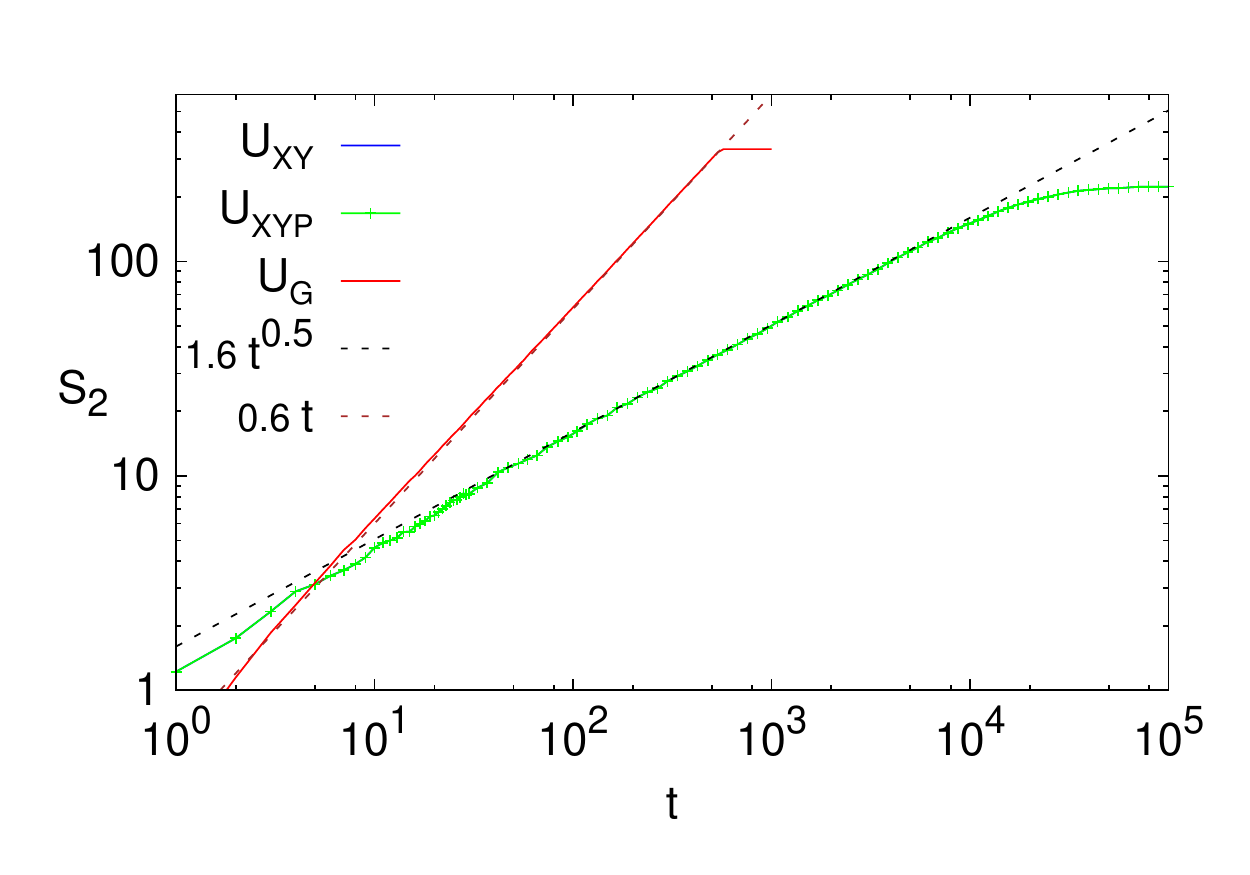}}
\caption{(Color online) Average $S_2$ for 1D random Clifford qubit system, middle-$\frac{1}{3}$ bipartition, and gates $\UXY$, $U_{\rm XYP}$ and $U_{\rm G}$. Data for magnetization conserving $\UXY$ and $U_{\rm XYP}$ almost overlap, both growing diffusively. All is for $L=1002$ and averaged over $100$ realizations.}
\label{fig:XYP}
\end{figure}

In the main text we used the XY gate in qubit ($q=2$) systems to demonstrate diffusive growth of $S_2$. Such gate is quadratic in fermionic operators. Here we show that using more complicated Clifford gates (which in particular are not quadratic), similarly as has been done for $q>2$, leads to similar results. In Fig.~\ref{fig:XYP} we show data for $S_2$ for three different types of Clifford circuits. One that uses only the XY gate (similar data as in Fig.~\ref{fig:plot123D}(a)), one with a gate $U_{\rm XYP}=\UXY \exp{[-\ii \frac{\pi}{4}(\sz{1}+\sz{2})]}$, and one with a gate $U_{\rm G}={\rm CNOT}_{12}\,{\rm H}_1 \exp{(\ii \frac{\pi}{4}\sz{2})}$ that does not conserve the magnetization.

\begin{figure}[t!]
\centerline{\includegraphics[width=.8\columnwidth]{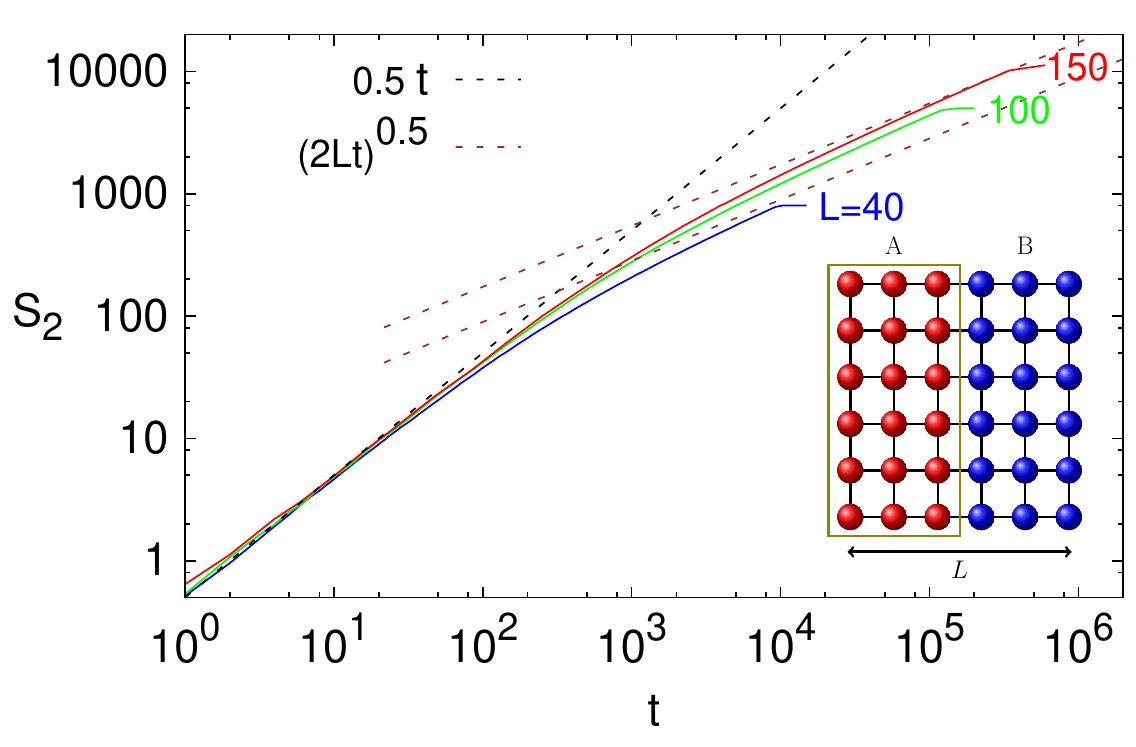}}
\caption{(Color online) Entanglement growth for 2D Clifford qubit system and the half-cut bipartition.}
\label{fig:plot2D}
\end{figure}

In Fig.~\ref{fig:plot2D} we show data for the same 2D diffusive qubit Clifford circuit utilizing $\UXY$ gate as in Fig.~\ref{fig:plot123D}b, but for a half-cut bipartition. We can see that the agreement with theoretical short-time as well as long-time prediction (Table~\ref{tab:summary}) is good. The short-time growth is $S_2 \approx 0.5 t$, where $0.5=\frac{l}{2L}$ with $l=L$, whereas the asymptotic growth goes into $S_2 \asymp \sqrt{2Lt}$ (no fitting parameters).

\section{Additional data for non-Clifford qutrit systems (spin $S=1$)}
\label{app:q3}
In the main text we demonstrated that the dynamics given by the Clifford qutrit gate $\U3$, which conserves both the non-trivial diagonal operators $Z_1+Z_2$ and $Z_1^2+Z_2^2$, results in a diffusive asymptotic growth of $S_2$ (Fig.~\ref{fig:plotd3}). Instead of the two ``Clifford''-basis diagonal operators $Z_j$ and $Z^2_j$ we can also use the language of spin $S=1$ particles: there the two non-trivial diagonal operators are $S^{\rm z}_j={\rm diag}(1,0,-1)$ and $\tS^{\rm z}_j=3(S^{\rm z}_j)^2-2\cdot \1_j={\rm diag}(1,-2,1)$. The gate $\U3$ of course also conserves those two, $\U3^\dagger (S^{\rm z}_1+S^{\rm z}_2)\U3=S^{\rm z}_1+S^{\rm z}_2$, $\U3^\dagger (\tS^{\rm z}_1+\tS^{\rm z}_2)\U3=\tS^{\rm z}_1+\tS^{\rm z}_2$. Note that $\U3$ is up-to phases equal to the spin-1 SWAP gate $U_{\rm SWAP}=\exp{(-\ii \frac{\pi}{2}[(\mathbf{S}_1\cdot \mathbf{S}_2)^2+(\mathbf{S}_1\cdot \mathbf{S}_2)+2\cdot\1])}$, $\U3 U_{\rm SWAP}^\dagger={\rm diag}(1,1,1,1,\omega,\omega^2,1,\omega^2,\omega)$.

We are going to show additional data for a number of spin-1 quantum circuits that are not of the Clifford type. We will always use a half-cut bipartition and the same initial state as used for Clifford circuits, that is a uniform superposition of all computational states, $\psi(0)\sim (\ket{1}+\ket{0}+\ket{\text{-}1})^{\otimes L}$. The shown $S_2$ is the average one over between $10^3$ (small $L$) and $20$ (for largest $L=21$) circuit realizations. The aim is to further shed light on the fact that we expect the asymptotic $S_2 \sim \sqrt{t}$ growth only if all diagonal operators are conserved and diffusive. For $q=3$ this means that both $S^{\rm z}$ and $\tS^{\rm z}$ should be conserved (alternatively, both $Z$ and $Z^2$).

\begin{figure}[t!]
\centerline{\includegraphics[width=.8\columnwidth]{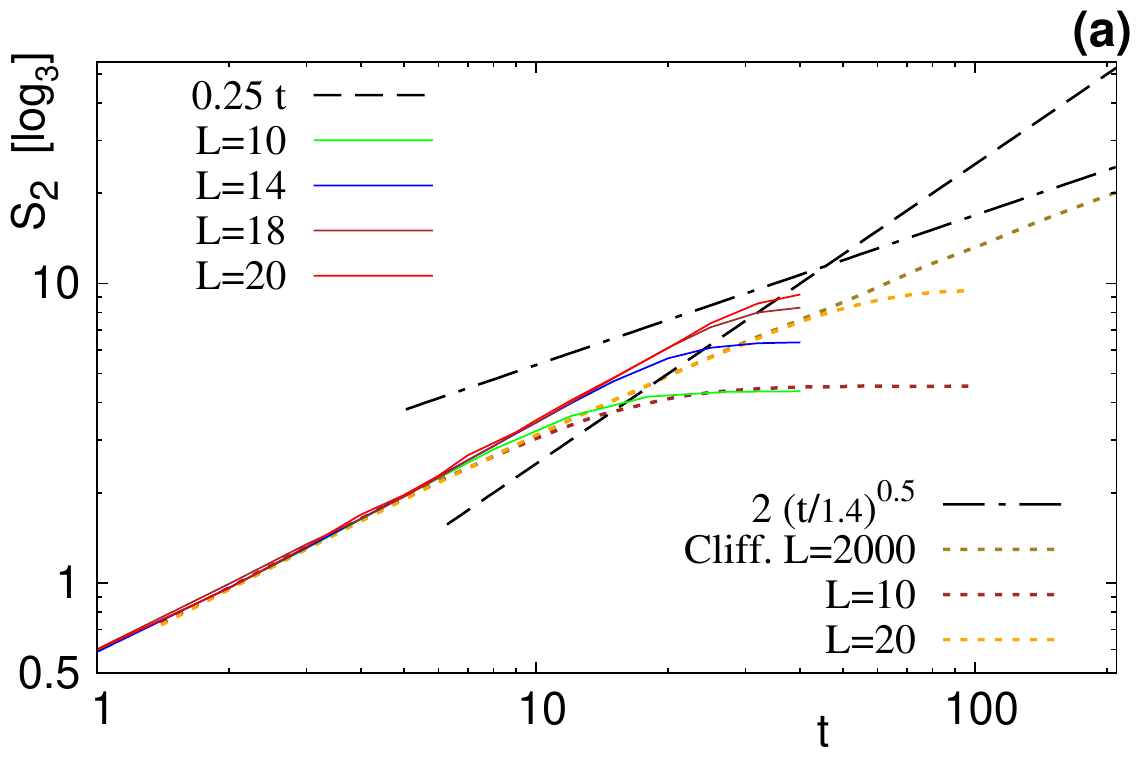}}
\centerline{\includegraphics[width=.8\columnwidth]{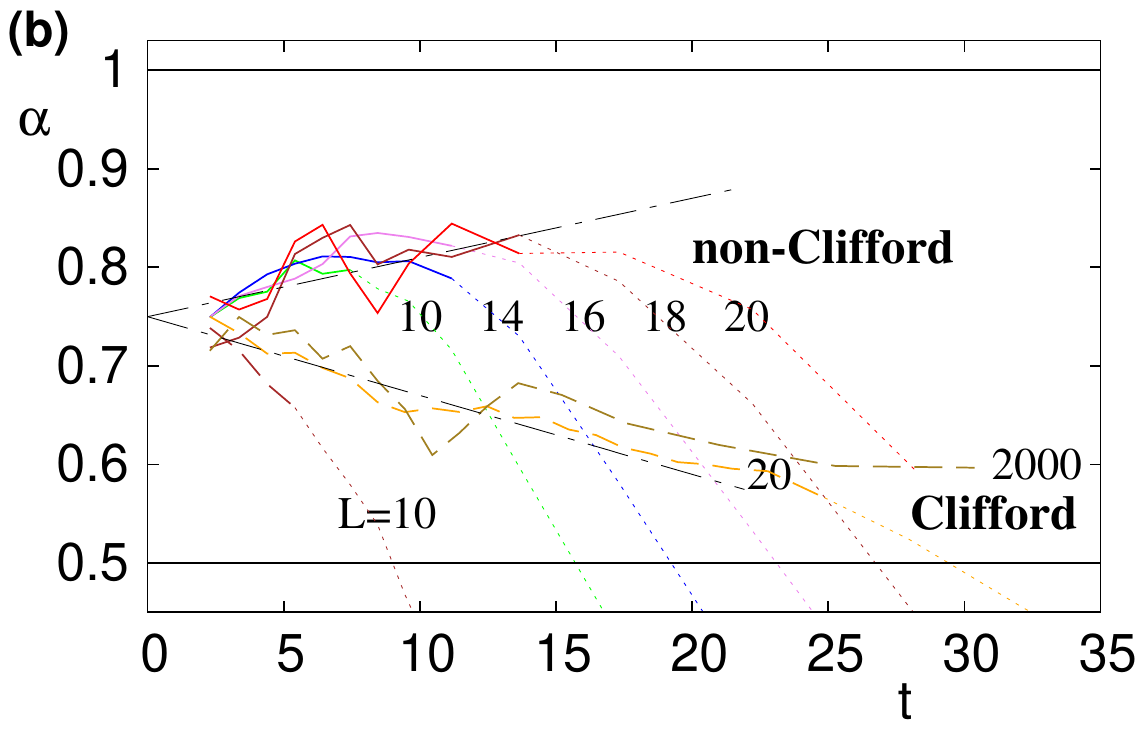}}
\caption{(Color online) (a) $S_2(t)$ for two spin-1 random circuits: a Clifford system with $\U3$ conserving both diagonal operators and resulting in diffusive growth (dotted curves for $L=10,20,2000$ sites; times are multiplied by $1.4$), and a non-Clifford $U_{\rm I}$ (full curves for $L=10,14,18,20$) that conserves only one diagonal operator and is conjectured to lead to the asymptotic linear growth. The logarithm in the definition of $S_2$ is here base-$3$. In (b) we plot a finite-time scaling exponent $\alpha$ (from $S_2 \sim t^\alpha$) for the same data (solid lines for $U_{\rm I}$, dashed for $\U3$; line-style is changed into a dotted curves for times when $S_2$ starts to approach a finite-$L$ saturation value). Chain lines suggest convergence with $L$.}
\label{fig:exact-Ucomp}
\end{figure}

We first check the evolution using a 2-site gate $U_{\rm I}$ that is a concatenation of the diffusive $\U3$ we already used in the main text and the isotropic gate, that is $U_{\rm I}=U_{\rm ISO}\U3$, where $U_{\rm ISO}=\exp{(-\ii \frac{\pi}{\sqrt{2}} \mathbf{S}_1\cdot\mathbf{S}_2)}$. The gate $U_{\rm ISO}$ is not a Clifford gate, and conserves $S^{\rm z}$, $U_{\rm ISO}^\dagger (S^{\rm z}_1+S^{\rm z}_2) U_{\rm ISO}=S^{\rm z}_1+S^{\rm z}_2$, but not $\tS^{\rm z}_1+\tS^{\rm z}_2$. The gate $U_{\rm I}$ therefore conserves only $S^{\rm z}_1+S^{\rm z}_2$, the dynamics of which is diffusive due to the spatial randomness (a gate is applied to a random n.n. bond). We therefore expect the asymptotic growth of $S_2$ to be linear for $U_{\rm I}$ despite diffusive total magnetization. The results are shown in Fig.~\ref{fig:exact-Ucomp}. We immediately have to remark that a drawback of non-Clifford circuits is that only very small systems can be simulated, and correspondingly the reachable times are far from the asymptotic ones. 
For comparison we also show data for the Clifford circuit with $\U3$ and $L=2000$ (the same data as in Fig.~\ref{fig:plotd3}), as well as for $L=10$ and $L=20$. Because the initial rates of the entropy production are different for $\U3$ and $U_{\rm I}$ we multiply the times for the $\U3$ data by $1.4$ so that the curves overlap at short times. We can see that up-to times $t \approx 6$ the two evolutions result in the same growth of $S_2$ (one could fit $S_2 \approx 0.59 t^{0.76}$); for instance, for $L=10$ qutrits it is hard to claim any difference between $\U3$ and $U_{\rm I}$. After $t \approx 10$ though deviations start to appear: the Clifford case $\U3$ that conserves both diagonal operators starts to converge to slower $S_2 \asymp \sqrt{t}$ growth, whereas $U_{\rm I}$ that conserves only the total $S^{\rm z}$ starts to grow faster. This is furthermore seen also in the dependence of the scaling exponent in $S_2 \sim t^\alpha$ on time. Due to taking a numerical derivative the data for $\alpha$ is much more noisy (particularly for $L=20$ where the ensemble size is 100), however one can nevertheless see a clear difference between the two cases (in-line with observation in Fig.~\ref{fig:exact-Ucomp}(a)). For the non-Clifford circuit $\alpha$ increases with time, while for the Clifford case it decreases. While from such short-time data it is impossible to make a definite claim about the asymptpotic growth, what we observe is compatible with the asymptotics $S_2 \sim t$ for $U_{\rm I}$. If the Clifford case, where we can simmulate large systems, is any indication of the required sizes necessary to reach the asymptotics, we can say that likely about 5 times larger systems would be required to really see the asymptotic linear growth for $U_{\rm I}$ (for the Clifford data in Fig.~\ref{fig:plotd3}, we can see that one converges to $S_2 \sim \sqrt{t}$ only at $t \approx 10^3$ where $S_2 \approx 10^2$).

\begin{figure}[t!]
\centerline{\includegraphics[width=.8\columnwidth]{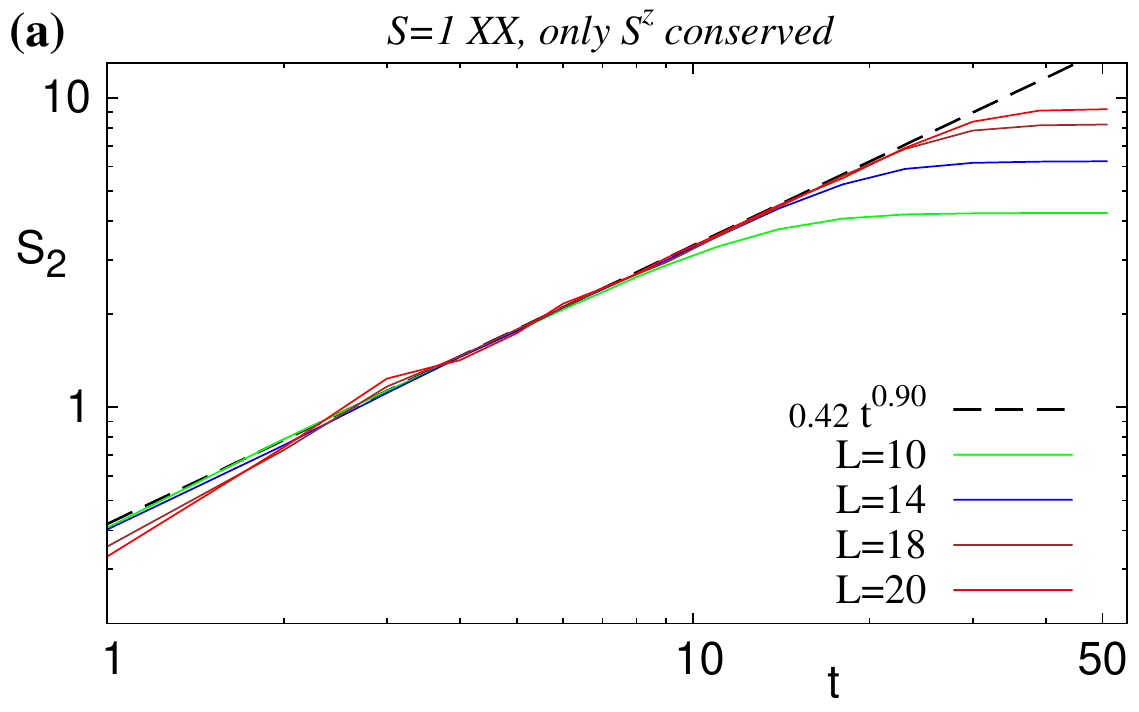}}
\centerline{\includegraphics[width=.8\columnwidth]{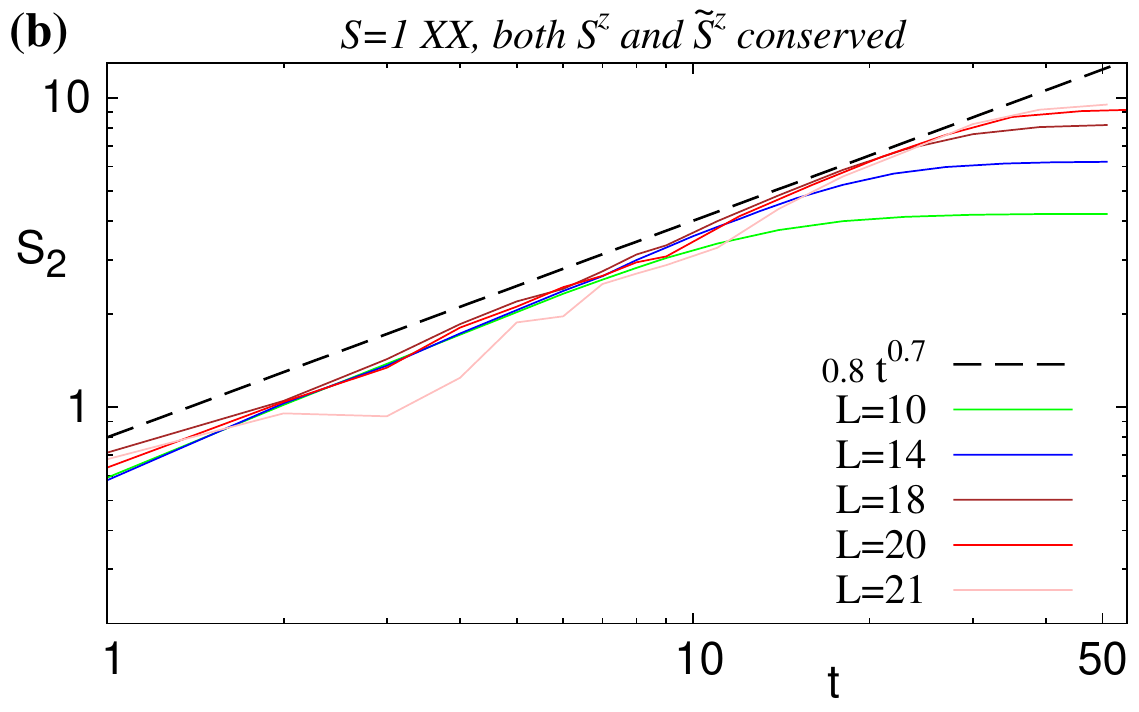}}
\centerline{\includegraphics[width=.8\columnwidth]{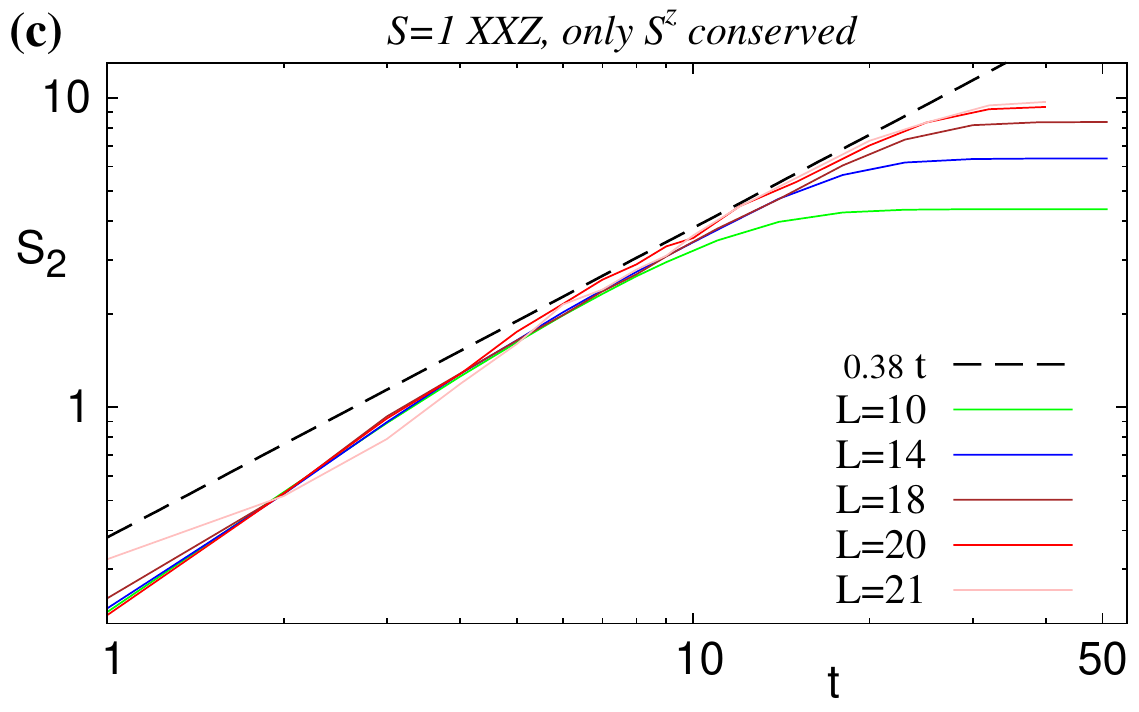}}
\caption{(Color online) $S_2(t)$ for three non-Clifford spin-1 random circuits. Top frame (a) is for $U_{\rm XX2}$, frame (b) for $U_{\rm XX}$, and frame (c) for $U_{\rm XXZ}$ (see text for details). The logarithm in the definition of $S_2$ is always base-$3$.}
\label{fig:exact}
\end{figure}

In Fig.~\ref{fig:exact} we show data for further non-Clifford random circuits. We show results for $U_{\rm XX2}=\exp{(-\ii \frac{\pi}{2\sqrt{2}} [S^{\rm x}_1 S^{\rm x}_2 + S^{\rm y}_1 S^{\rm y}_2])}$ that conserves only $S^{\rm z}_1+S^{\rm z}_2$, but not $\tS^{\rm z}_1+\tS^{\rm z}_2$. Data in frame (a) is compatible with the linear asymptotic growth. If we on the other hand change the gate to $U_{\rm XX}=\exp{(-\ii \frac{\pi}{\sqrt{2}} [S^{\rm x}_1 S^{\rm x}_2 + S^{\rm y}_1 S^{\rm y}_2])}$, which conserves both $S^{\rm z}_1+S^{\rm z}_2$ and $\tS^{\rm z}_1+\tS^{\rm z}_2$, we see that the growth is much slower, like $S_2 \sim t^{0.7}$ at short times. While it is hard to make any asymptotic claims about $S_2 \sim \sqrt{t}$ based on such numerics (exact numerics for larger systems gets hampered by memory requirements; the Hilbert space size for $L=21$ qutrits is about the same as for $\approx 33$ qubits), what is very distinct is that the exponent is very different in (a) and (b) despite a very similar 2-site gate; the only difference between the two is the conservation of $\tS^{\rm z}_1+\tS^{\rm z}_2$. Finally, as a third example we show the anisotropic XXZ-like gate $U_{\rm XXZ}=\exp{(-\ii \frac{\pi}{3}[S^{\rm x}_1 S^{\rm x}_2 + S^{\rm y}_1 S^{\rm y}_2+1.5 S^{\rm z}_1 S^{\rm z}_2])}$ that again conserves only the total magnetization, and therefore one has $S_2 \sim t$ visible already at short times.

\section{Data for ladder systems ($q=4$)}
\label{app:q4}
Here we show further data supporting the claim that for ladders the growth of $S_2$ is generically linear in time. We simulate Clifford ladders with a large number of rungs $L=10000$ using different 2-site gates. On legs, upper or lower, we apply either the already seen $\UXY=\exp{(-\ii \frac{\pi}{4}(\sx{j}\sx{k}+\sy{j}\sy{k}))}$, or $U_{\rm G}={\rm CNOT}_{12}\,{\rm H}_1 \exp{(\ii \frac{\pi}{4}\sz{2})}$. On the rungs we use either $U_{\rm ZZ}=\ii \exp{(-\ii \frac{\pi}{2}\sz{1}\tau^{\rm z}_2)}$, or $U_{\rm Sxy}=\exp{(-\ii \frac{\pi}{4}\sz{1})} \exp{(-\ii \frac{\pi}{4}\tau^{\rm y}_2)}\exp{(-\ii \frac{\pi}{4}\tau^{\rm z}_2)}$. The protocol is always the same: at each step we apply one of the leg gates on a random bond on either the upper or the lower leg, and a rung gate on an independent random rung. The type of the gate applied on rungs as well as on the upper and lower leg is held fixed, so that one can get $8$ different protocols out of the 4 mentioned gates.

\begin{figure}[t!]
\centerline{\includegraphics[width=.8\columnwidth]{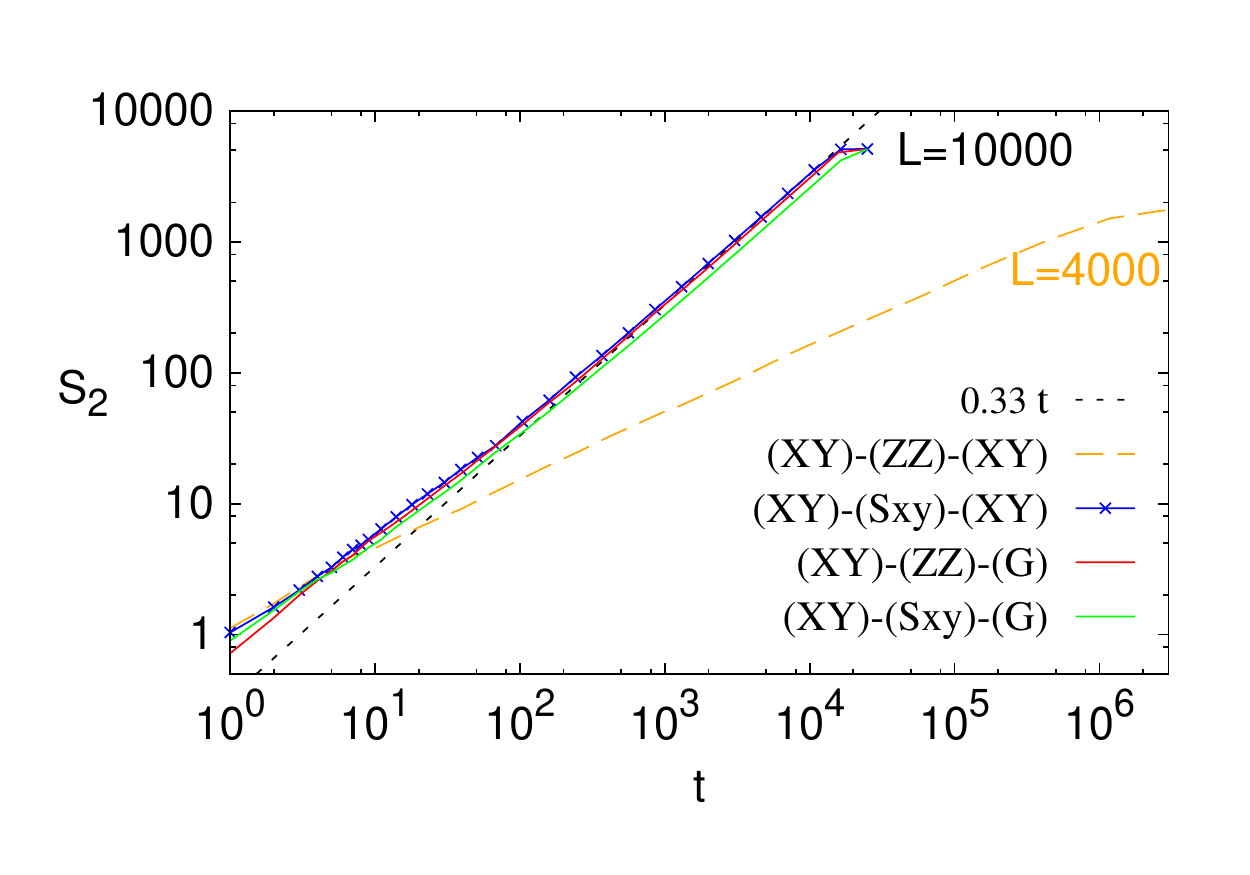}}
\caption{(Color online) Entanglement growth for different ladder systems ($q=4$) and the half-cut bipartition. Notation $(A)-(B)-(C)$ denotes Clifford evolution with the gate $U_A$ on the upper leg, $U_B$ on the rung, and $U_C$ on the lower leg. For instance, the orange data $(XY)-(ZZ)-(XY)$ (as well as the red one) is the same as the one already shown in Fig.~\ref{fig:plotd3}.}
\label{fig:ladders}
\end{figure}
The gate $\UXY$ conserves the total magnetization on the respective leg on which it acts, while $U_{\rm G}$ does not. Namely, $U_{\rm G}^\dagger \1_j\sz{k} U_{\rm G}= \sx{j}\sz{k}$, and $U_{\rm G}^\dagger \sz{j} \1_k U_{\rm G}= \sx{j}\1_k$, so that one has $U_{\rm G}^\dagger (\sz{j}+\sz{k}) U_{\rm G}=\sx{j}(\1_k+\sz{k})$. The rung gate $U_{\rm ZZ}$ does not break conservation of $\sz{1}+\sz{2}$, nor of $\tau^{\rm z}_1+\tau^{\rm z}_2$ because one has $U_{\rm ZZ}^\dagger \sz{1}\1_2 U_{\rm ZZ}= \sz{1}\1_2$, and $U_{\rm ZZ}^\dagger \1_1 \tau^{\rm z}_2 U_{\rm ZZ} = \1_1 \tau^{\rm z}_2$ (as well as $U_{\rm ZZ}^\dagger \sx{1}\1_2 U_{\rm ZZ}= -\sx{1}\1_2$, $U_{\rm ZZ}^\dagger \1_1 \tau^{\rm x}_2 U_{\rm ZZ}= -\1_1\tau^{\rm x}_2$). The gate $U_{\rm ZZ}$ though does introduces non-trivial phases $\pm 1$ in the dynamics of Pauli $x$ and $y$ matrices. The gate $U_{\rm Sxy}$ on the other hand preserves conservation of magnetization only on the upper leg, $U_{\rm Sxy}^\dagger \sz{1}\1_2 U_{\rm Sxy}= \sz{1}\1_2$, $U_{\rm Sxy}^\dagger \sx{1}\1_2 U_{\rm Sxy}= -\sy{1}\1_2$, while it breaks conservation on the lower leg, $U_{\rm Sxy}^\dagger \1_1 \tau^{\rm z}_2 U_{\rm Sxy}= \1_1 \tau^{\rm y}_2$, $U_{\rm Sxy}^\dagger \1_1 \tau^{\rm x}_2 U_{\rm Sxy}= \1_1 \tau^{\rm z}_2$.

In Fig.~\ref{fig:ladders} we show results of numerical simulation for different protocols. Taking the $(XY)-(ZZ)-(XY)$ protocol where the leg gates $\UXY$ as well as the rung gates $U_{\rm ZZ}$ conserve magnetization on the upper and the lower leg, dynamics of all diagonal operators is diffusive and one has $S_2 \asymp \sqrt{t}$. The same data for $L=4000$ has been already shown in Fig.~\ref{fig:plotd3}. We can break conservation of magnetization on the lower leg by using $U_{\rm G}$, which as we can see results in the asymptotic growth $S_2 \asymp t$ (red curve in Fig.~\ref{fig:ladders}), the same data as in Fig.~\ref{fig:plotd3}). We can however break the conservation on the lower leg also by changing the rung gate to $U_{\rm Sxy}$. This is illustrated by the protocol $(XY)-(Sxy)-(XY)$, which again results in $S_2 \asymp t$. Note that here the dynamics along the two rungs is purely diffusive -- the gate $\UXY$ is used on both legs -- it is only the non-trivial rung dynamics that breaks one $U(1)$ symmetry and causes the asymptotic linear growth of $S_2$ (similar result would be obtained also if at each step of the protocol the $\UXY$ gate would be applied simultaneously to a pair of upper- and lower-leg bonds forming a local plaquete with two rungs). Finally, using conservation breaking $U_{\rm G}$ on the lower leg as well as the rung $U_{\rm Sxy}$, one again has $S_2 \asymp t$.

\end{document}